# Experimental observation of intrinsic light localization in photonic icosahedral quasicrystals

*Artem D. Sinelnik, Ivan I. Shishkin, Xiaochang Yu, Kirill B. Samusev, Pavel A. Belov, Mikhail F. Limonov, Pavel Ginzburg, Mikhail V. Rybin\**

((Optional Dedication))

A. D. Sinelnik, Dr. I. I. Shishkin, X. Yu, Dr. K. B. Samusev, Prof. P. A. Belov, Prof. M. F. Limonov, Prof. P. Ginzburg, Prof. M. V. Rybin.
ITMO University, St Petersburg 197101, Russia

X. Yu
Huazhong University of Science and Technology, Wuhan, 430074, China

Dr. I. I. Shishkin, Prof. P. Ginzburg
Department of Electrical Engineering, Tel Aviv University, Ramat Aviv, Tel Aviv 69978, Israel

Dr. K. B. Samusev, Prof. M. F. Limonov, Prof. M. V. Rybin.
Ioffe Institute, St Petersburg 194021, Russia
E-mail: m.rybin@metalab.ifmo.ru

One of the most intriguing problems of light transport in solids is the localization that has been observed in various disordered photonic structures[1-11]. The light localization in defect-free icosahedral quasicrystals has recently been predicted theoretically without experimental verification[10]. Here we report on the fabrication of submicron-size dielectric icosahedral quasicrystals and demonstrate the results of detailed studies of the photonic properties of these structures. Here, we present the first direct experimental observation of intrinsic light localization in defect-free quasicrystals. This result was obtained in time-resolved measurements at different laser wavelengths in the visible. We linked localization with the aperiodicity of the icosahedral structure, which led to uncompensated scattering of light from an individual structural element over the entire sphere, providing multiple scattering inside the sample and, as a result, the intrinsic localization of light.

The concept of quasicrystal as an aperiodic structure with long-range ordering was introduced in physics by D. Levine and P. J. Steinhardt[12-14]. Quasicrystals can be positioned between crystalline and amorphous materials as they possess a nontrivial symbiosis of the photonic properties of these two well-defined condensed-matter states. A key property of photonic crystals is the existence of energy pseudogaps that arise during multiple scattering of photons by lattices of periodically varying refractive indices. For certain photonic structures, pseudogaps merge into a complete band gap in three dimensions leading to the localization of light in a cavity mode[15,16]. In the case of disordered structures[17], one of the most intriguing properties is the effect of Anderson localization of light[18], the phenomenon that has been observed in a variety of structures of different dimensions[19]. Historically, the study of Anderson localization has focused on disordered substances, although the possibility of observing localization in perfect quasicrystals has also been discussed. However, the effect was observed only in 2D quasicrystalline structures in the presence of strong disorder[8] or under the action of nonlinearity[20]. Nevertheless, the existence of a clearly defined photonic band structure in 3D icosahedral quasicrystals[21], the observation of Bragg diffraction[22], the multiple scattering of light within the structure[23], and laser generation from dye-doped samples[24] were promising signs for the possibility of an experimental observation of intrinsic light localization in defect-free 3D quasicrystals **Figure 1**. Additional evidence is a recent report in which intrinsic localization was theoretically found in icosahedral quasicrystals[10]. Although the Ioffe-Regel criterion[25] for localization in disordered materials (the wavelength $\lambda$ becomes comparable to the transport mean free path $l^*$, that is $kl^* \leq 1$, where $k=2\pi/\lambda$ is the wave vector) is not fulfilled in icosahedral quasicrystals, the authors argued that band flattening at high frequencies of the calculated photonic band structure corresponds to a slower group velocity and a decrease in the scattering mean free path which increases the possibility of wave localization in terms of the Ioffe-Regel condition[10]. Here, to the best of our knowledge, we present the first experimental observation of intrinsic light localization in defect-free 3D quasicrystalline material. In addition, we studied Bragg diffraction in the visible region on the same samples uncovering their regular structure.

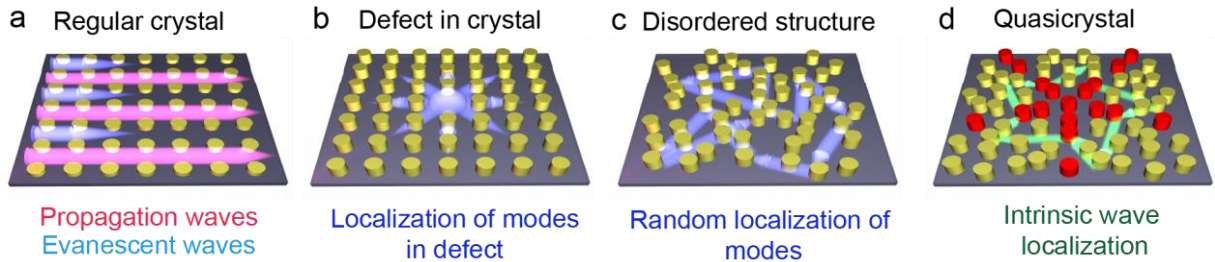

**Figure 1 Schematic representation of light transport regimes in various photonic structures.** a) Regular crystal. In a regular periodic structure, light propagates according to Bragg's conditions. In such a structure, there are propagating waves and evanescent waves. b) Defect in crystal. In a defective structure, light will be localized on this defect. As a result, there will exist defect-associated localized modes. c) Disordered structure. In a disordered medium, light is localized randomly due to the disorder. In this structure, randomly localized modes or Anderson localization are obtained. d) Quasicrystal. In a quasicrystal, light is localized, however, this is no longer randomization, but intrinsic localization in an ordered structure.

**Experimental Section**

*Sample Fabrication.* To fabricate experimental samples, we first generated a computer model of an icosahedral quasicrystalline structure in accordance with the substitution rules[10]. These quasicrystals had icosahedral symmetry with fifteen $C_{2v}$, ten $C_{3v}$ and six $C_{5v}$ axes[21], which led to the absence of periodicity, despite the fact that the structures had perfect ordering and regularity. Our samples consisted of six radial layers of icosahedral quasicrystalline unit cells (a total of 8112 connecting rods). We considered the rod length to be a quasicrystalline lattice constant, which was set at 3 μm. As a reference, we generated a model of a woodpile photonic crystal (with a lattice constant of 3 μm), with a shape limited to a sphere of the corresponding radius **Figure 2 (a)**. The samples were prepared by direct laser writing technique[26,27] using a hybrid organic-inorganic material based on zirconium propoxide with a refractive index of about $n = 1.52$ (see Supporting Information).

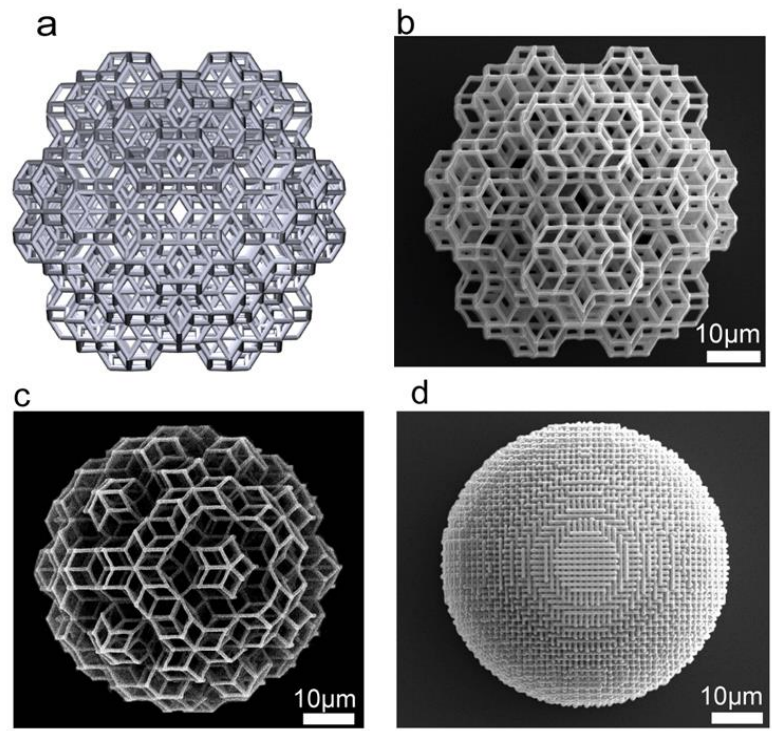

**Figure 2 Computer 3D model and SEM images of the fabricated structures**. a), 3D computer model of the icosahedral quasicrystal oriented along a 2-fold symmetry axis. b, c,) SEM images of the fabricated icosahedral quasicrystal with orientation along a 2-fold symmetry axis (b) and along a 5-fold symmetry axis (c). d), SEM image of a fabricated woodpile photonic crystal whose shape is limited to a sphere. Thus, both structures had a relatively large total radii of about 25 μm (i.e. from 34 to 65 wavelengths in the visible spectrum), which allowed to search for the intrinsic localization of light in our samples. Based on the SEM images, the diameter of the rods was estimated at 400 nm.

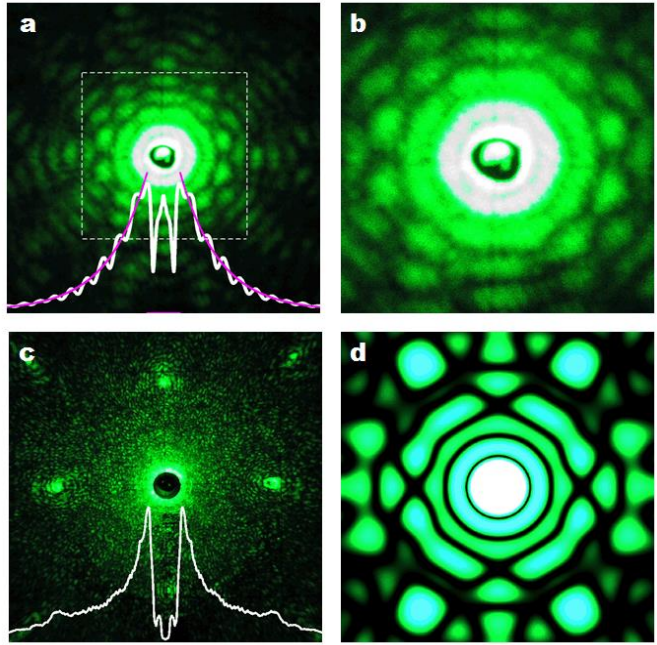

**Figure 3 Experimental and calculated diffraction patterns**. a) Experimental diffraction pattern of the icosahedral quasicrystal oriented along a 2-fold symmetry axis. The white curves show the averaged intensity profile, taken over the entire screen area and normalized to the number of pixels on each circle. The red curve is the exponentially decaying fit of the white curve. b) The central part of the experimental diffraction pattern, indicated by the dotted line on the panel (a). c) Experimental diffraction pattern of the woodpile crystal. The white curves show the averaged intensity profile, taken over the entire screen area and normalized to the number of pixels on each circle. d) The calculated diffraction pattern of the icosahedral quasicrystal for the region of a flat screen corresponding to the panel (b). Laser wavelength $\lambda=532$ nm.

*Observation Bragg diffraction.* To analyze the crystalline photonic properties of quasicrystals, we employed far-field measurements of Bragg diffraction patterns. A light beam from a 532 nm laser was used to illuminate the samples. Diffraction patterns in the forward scattering geometry were observed with the naked eye on a flat semitransparent screen and were recorded by a digital camera placed behind the screen. **Figure 3** summarizes the measured and calculated Bragg diffraction patterns. For the woodpile crystal, the diffraction pattern demonstrated the $C_{4v}$ symmetry and consisted of two components: Bragg reflections associated with the crystal structure and speckle-type background due to scattering on individual rods Figure 3 (c). The second scattering component was determined by eigenmodes of finite-length rods[28], and for some values of the aspect ratio, supercavity modes appeared in the scattering spectra[29,30]. Due to the different rod lengths, the scattering was random which led to a speckle pattern. The ratio of the length of the rods (from 10 to 50 μm) and the laser wavelength (~0.5 μm) determined the Fraunhofer scattering regime with a narrow lobe Figure 3 (c).

Compared with the woodpile structure, the quasicrystalline diffraction appears fundamentally different Figure 3. Firstly, the speckle component is not observed Figure 3 (a) due to the fact that all the rods forming the icosahedral have the same length. In periodic photonic systems the diffraction spots and transmission bandgaps are complementary since the energy conservation and they both have the same origin, that is the Bragg diffraction (for example, see reference 31). Although quasicrystals do not have any fragments with a sufficiently long periodic structure diffraction maxima are still connected to pseudogaps in the frequency spectra because the diffraction pattern has a close relation to the reciprocal (Fourier) space[32]. The pronounced patterns in Figure 3 (a,b) of unconventional Bragg diffraction indicate the existence of multiple photonic pseudogaps in our quasicrystals. The Bragg reflexes turned out to be very wide; moreover, they overlapped for many directions in space, Figure 3 (a,b). Icosahedral quasicrystals have a higher point group symmetry than ordinary crystals and the measured Brillouin zone is close to spherical[20]. These are optimal conditions for the formation of a complete photonic band gap from multiple pseudogaps and for trapping the light.

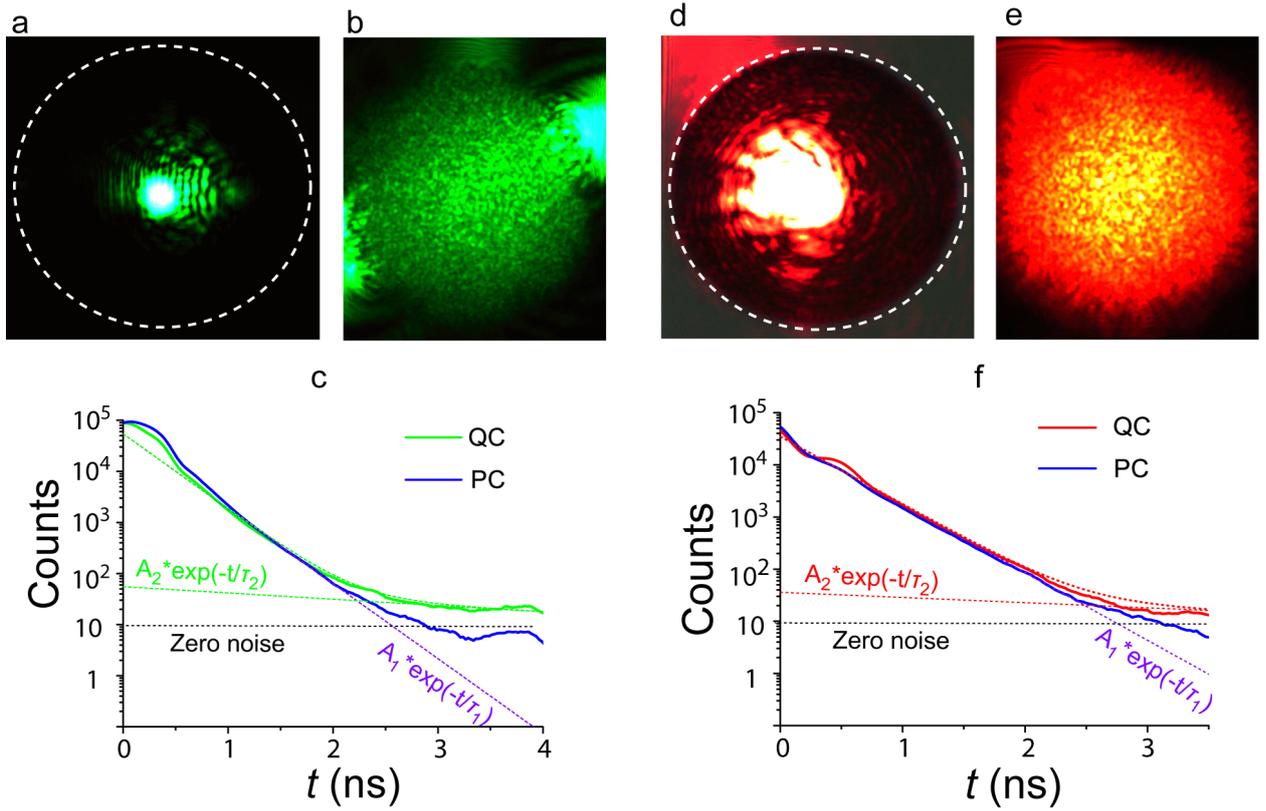

**Figure 4 Localization of light in quasicrystals.** a,b) Image in the optical microscope showing a propagating beam and a weak ripple structure around, arising from Fraunhofer scattering for two wavelengths of 530 nm and 630 nm, respectively. The sample boundaries are shown by a white dashed circle. b, e,) Optical microscope images of icosahedral quasicrystal which is completely filled with scattered light due to intrinsic localization for two wavelengths. c, f,) Experimental and fitting curves showing the distribution of light intensity elapsed through the photonic crystal (PC) and quasicrystal (QC) (along a 2-fold symmetry axis) as a function of time for two wavelengths. A1=99%, A2=1% counts for both wavelengths. $\tau_1$= 0.29 ns and $\tau_2$=3.3ns for wavelength 530 nm (c). $\tau_1$= 0.3 ns and $\tau_2$=4.3 ns for wavelength 630 nm (f).

*Observation of intrinsic light localization.* Recently, localization in defect-free photonic quasicrystals has been predicted theoretically for the case of two- and three-dimensional structures[10,33] using finite-difference time-domain methods. We investigated the light transport properties of icosahedral quasicrystals and woodpile crystals of the same spherical shape and size using optical microscopy and analysis of the picosecond pulse propagation delay. In our experiments, two wavelengths of 530 nm and 630 nm, generated by a supercontinuum laser source, were used to study the intrinsic light localization properties of our samples (see Figure S2). **Figure 4** shows images of the woodpile and icosahedral samples captured by an optical microscope camera. Figure 4a clearly demonstrates that there is only a ballistic component and a weak ripple structure (successive bright and dark stripes) around it due to Fraunhofer diffraction with a narrow lobe Figure 3 (c). The optical images were recorded by a digital camera (see Figure S2). However, the rest of the spherical sample does not reveal any light scattering. In the case of icosahedral quasicrystals, the optical microscope images differ dramatically. Figures 4 (b,e) show that the entire volume of the quasicrystal is completely filled with scattered light.

Periodicity of a system enables the wave vector as a quantum number classifying all modes to be either propagating (real values) or evanescent (complex values) waves related to bandgap frequencies. The physical meaning of this is the following. Multiple scattering events in the periodic structure result to the destructive interference in all directions except the forward one. The latter is possible due to the constructive interference for the propagating waves and destructive one for the bandgap frequency. The lack of translation symmetry reduces the prefect constructive and destructive interference condition from the phase of re-scattered waves, which lead for the light diffusion to occur (compare Figure 4 (a,c) and 4 (b,d) ). Unlike disordered media the regular structure of quasicrystals keeps the coherence of wave at a large distance allowing wave localization for relatively high values of $kl^*$.

One method for studying the intrinsic light localization in an experiment is to exploit time-domain measurements. Light transport properties of our samples were studied by using a transmission light microscope setup enhanced with time-correlated single photon counting (TCSPC) capability. Figures 4 (c,f) present the transmitted intensity as a function of time at wavelengths of 530 nm and 630 nm through quasicrystal and woodpile samples. We also measured reference curves for a bare substrate without samples on its surface. In the case of the woodpile crystal, the intensity exhibited an almost exponential decay over 3 ns until it fell below the zero-level noise of our equipment. The decay curve best fitted with a one-exponential decay model with a decay time of $\tau = 0.29$ ns for both 530 nm and 630 nm wavelengths. The curves obtained for the empty substrate gave the same values of decay time, which meant that this decay time could be attributed to the intrinsic response of the measurement setup rather than a shorter delay time required to pass a pulse through the woodpile photonic crystal operating in the ballistic regime. The intensity decay curves Figures 4 (c,f) of the picosecond pulse transmission through the quasicrystal along a 2-fold symmetry axis were best fitted with a dual-exponential decay model determining the values of $\tau_1 = 0.29$ ns for the wavelength 530 nm and $\tau_1 = 0.30$ ns for 630 nm (with a fractional amplitude of $A_1$=99% for 530 nm and 630 nm both). These decay times corresponded to the intrinsic instrument response. The second exponent was related to the process with the decay time of $\tau_2 = 3.3$ ns (530 nm) and $\tau_2 = 4.3$ ns (630 nm) with the fractional amplitude $A_2$=1%). The values of $\tau_2$ were longer by an order of magnitude than $\tau_1$. The results of fitting the decay curve to the dual-exponential model are shown in Figures 4 (c,f). Double-exponential fit of the data showed that the pulse transmission was associated with fast and slow processes. The first (fast) process was ballistic light transport, corresponding to the case when almost all the photons passed through the structure without a delay. Owing to the small sample thickness (< 50μm), the time delay in photon propagation through the structure could not be detected with the TCSPC setup used. However, we observed the second (slow) process in icosahedral quasicrystals, which is related to the intrinsic localization of light in regular structures. Similar results were obtained at the picosecond pulse transmission through the quasicrystal along a 5-fold symmetry axis (see Figure S3). We compared these results with the decay curves recorded for the woodpile photonic crystal fabricated by the same technological process and found that the latter did not contain a slow tail Figures 4 (c,d).

Let us discuss a difference between Anderson localization and intrinsic localization in quasicrystals reported here. The localization requires coherence between multiple scattering events. We remind that disordered systems with a high enough turbidity enable the transition to the Anderson localization regime and in theory the Ioffe-Regel condition describes the competing of wavelength and mean free path. However, this criterion is not rigorous. In particular, the Anderson localization was observed in $TiO_2$ powders having turbidity $kl^* = 4.5$ almost five times beyond the limit[9]. On the other hand, regular systems having pseudogaps necessitates a reinterpretation of the standard Ioffe-Regel criterion[15]. The distinct spots in optical diffraction patterns (Figure 3) uncover that quasicrystals allow strong interference effects in the superwavelength scales ($l \gg \lambda$). Thus, the standard Ioffe-Regel $kl^* \leq 1$ limit is not applicable for cases of localization in well-ordered though non-periodic systems.

Quasicrystals and woodpile samples were fabricated using the same technology and the same material, thus material absorption cannot cause the observed pulse delay. Therefore, all deviations from classical diffusion are direct evidence of the intrinsic light localization in 3D defect-free quasicrystals. The main reason for localization is the aperiodicity of the icosahedral structure that breaks the perfect Bragg condition of long-range periodic photonic crystals. As a result, light is scattered from individual rods into the entire sphere (see Figure S4), and not only along Bragg directions, as in the crystal lattice. This leads to multiple scattering of light inside the quasicrystals, as previously reported based on analysis of experimental data[23]. The first consequence is the glow of the entire sample Figures 4 (b,e), which is absent in the case of periodic woodpile Figures 4 (a,d). The second and most important consequence is a decrease in the mean free path and the intrinsic localization of light. The aperiodicity of quasicrystals turned out to be a mechanism similar to disorder in Anderson localisation[8]. Our results will fill the void in the field of light scattering between ordered and disordered structures and paves the way for a variety of applications in optics, from lasing[34,35] and sensing[36] to telecommunications[37] using defect-free structures which support the intrinsic photonic wave localization.